\documentstyle[11pt]{article}
\begin{document}
\begin{flushright}
\renewcommand{\textfraction}{0}
May 25th 1994\\
hep-th/9405172\\
\end{flushright}


\begin{center}
{\LARGE {\bf Event Symmetric Open String Field Theory } }
\end{center}
\begin{center}
{\Large Phil Gibbs} \\
\end{center}

\begin{abstract}
Field Theory on Event Symmetric space-time is constructed using the gauge
group of discrete open strings. Models with invariant actions
can be viewed as natural extensions of Matrix Models. The objective is to
find a fundamental non-perturbative pre-theory for superstrings.
\end{abstract}

\section{Introduction}

\subsection{String Theories}

Despite the lack of experimental data above the Electro-Weak energy scale, the
search for unified theories of particle physics beyond the standard model has
yielded many mathematical results based purely on constraints of high symmetry,
renormalisability and cancellation of anomalies. In particular, space-time
supersymmetry \cite{Wess,Zumino(1974)} has been found to improve perturbative
behaviour and to bring the gravitational force into particle physics. One
ambitious but popular line of research is superstring theory
\cite{Green,Schwarz(1981a,b,1982)}. Renormalisable string models were
originally constructed in perturbative form but were found to be incomplete in
the sense that the perturbative series were not Borel summable
\cite{Gross,Perival(1988)}.

More recently there have been various attempts to formulate Superstring
Field Theories non-perturbatively (see \cite{Kaku(1991)}). The use of the
Universal String Group to represent the gauge symmetry and formulate a
covariant action is an elegant approach but no non-perturbative formulation
has yet been found from which the true vacuum state can be derived.

A successful theory of Quantum Gravity should describe physics at the
Planck scale \cite{Planck(1899)}. It is possible that there is a phase
transition in string theories at their Hagedorn temperature
\cite{Hagedorn(1968)}. It has been speculated that above
this temperature there are fewer degrees of freedom and a restoration of a much
larger symmetry \cite{Atick,Witten(1988);Gross,Mende(1987,1988);Gross(1988)}.
Some interpret this as an indication that in the hot phase string theory is a
pure Topological Quantum Field Theory with one degree of symmetry for
each degree of freedom.

There is another non-perturbative approach to string theories which
gives important insights. Random Matrix Models in the large $N$ double
scaling limit can be shown to be equivalent to 2 dimensional gravity, or
equally, string theory in zero dimensional target space. The models can be
extended to a one dimensional target space but not to critical dimension where
string theory is perturbatively finite.

One interpretation of the present state of string theories is that it
lacks a geometric foundation and that this is an obstacle to finding
its most natural formulation. It is possible that our concept of
space-time will have to be generalised to some form of stringy space.
Perhaps such space-time must be dynamical and capable of undergoing
topological or dimensional changes \cite{Antoniadis,Ferrara,Kounnas(1994)}.

\subsection{Event Symmetric Discrete Models}

In the light of what quantum gravity seems to have to say about the
structure of space-time on small scales Wheeler and others have
speculated that a pre-geometry theory is needed \cite{Wheeler(1984)}. In such a
theory the space-time continuum would not be part of the fundamental
formulation but would arise as a consequence of dynamics. Pre-geometry may
have to have a discrete formulation.

The Event Symmetric approach is one such discrete method \cite{Gibbs(1994)}.
The exact nature of space-time in this scheme will only become apparent in the
solution. Even the number of space-time dimensions is not set by the
formulation and must by a dynamic result. It is possible that space-time will
preserve a discrete nature at very small length scales. The objective is to
find a statistical system which reproduces a unified formulation of known and
hypothesised symmetries in physics and then worry about states, observables
and causality later.

We seek to formulate a lattice theory in which diffeomorphism invariance
takes a natural and explicit discrete form. At first glance it would seem that
only translational invariance can be adequately represented in a discrete form
on a regular lattice but this overlooks the most natural generalisation of
diffeomorphism invariance in a discrete system. Diffeomorphism invariance
requires that the action should be symmetric under any differentiable
transformation of coordinates on a $D$ dimensional manifold. This is
represented by the diffeomorphism group $diff(M_D)$. On a discrete
space we  could demand that the action is symmetric under any permutation
of the discrete space-time events ignoring continuity altogether. Generally we
will use the term {\it Event Symmetric} whenever an action has a symmetry
given by the Symmetric Group $S(\infty)$ over an infinite number of discrete
``events'' (or some larger group of which it is a sub-group).

More precisely the symmetric group is the group of all possible 1-1 mappings
on a set of events. The cardinallity of events on a manifold of any number
of dimensions is $\aleph_1$. The number of dimensions of the manifold is lost
in an event symmetric model since the symmetric groups for two sets of equal
cardinallity are isomorphic.

Event symmetry is larger than the diffeomorphism invariance of continuum
space-time.
\begin{equation}
                 diff(M_D) \subset S(\aleph_1)
\end{equation}
If a continuum is to be restored there must be a spontaneous mechanism of
symmetry breaking in which event symmetry is replaced by a residual
diffeomorphism invariance. The mechanism will determine the number of
dimensions of space. It is possible that a model could have several phases
with different numbers of dimensions and may also have unbroken event-symmetric
phases.

It is unlikely that there would be any way to distinguish a space-time with an
uncountable number of events from space-time with a dense covering of a
countable number of events so it is acceptable to consider models in which the
events can be labelled with positive integers. The symmetry group $S(\aleph_1)$
is replaced with $S(\aleph_0)$. In practice it may be necessary to regularise
to a finite number of events $N$ with an $S(N)$ symmetry and take the large $N$
limit while scaling parameters of the model as functions of $N$.

\section{Event Symmetric String Models}

The fact that a large number of degrees of freedom are perhaps required to
produce event symmetry breaking suggests that string theories might
provide answers. The most natural place to start is with
string groups \cite{Kaku(1988)}. Such string groups are considered to
be the gauge groups of string theory and have been used to formulate both
open \cite{Witten(1986)} and closed \cite{Zwiebach(1992)} string field
theories.

These groups can also be formulated topologically without reference to a
space-time background and can be naturally placed in an event symmetric
setting.

\subsection{Discrete Open Strings}

A basis for a discrete open string vector space is defined by the set of
open ended oriented strings through an event symmetric space of $N$ events.
E.g. a possible basis element might be written,
\begin{equation}
              C = (1,4,3,1,7)
\end {equation}
Note that a string is allowed to cross itself. In the example the string
passes through the event $1$ twice. A string must be at least two events
long. A null string passing through zero events or Strings
passing through just one event could be included but are not needed. The
order in which the string passes through the events is significant e.g.
\begin{equation}
                   (1,4,3) \neq (1,3,4)
\end {equation}
but the order in which the events themselves have been numbered is
irrelevant since the models are to be event symmetric.
A complete set of field variables in an event symmetric string model would
be an element of this vector space which could be written
\begin{equation}
                     \Phi = \sum \Phi_C C
\end {equation}
This can define either a real or complex vector space.
There is also a scalar product
which can be defined. The scalar product is equal to $-1$ for two base strings
if one is the reverse of the other and zero otherwise. e.g.
\begin{equation}
                    (1,4,3,1,7) \bullet (7,1,3,4,1) = -1\\
		    (1,4,3,1,7) \bullet (3,4,4) = 0
\end{equation}
The scalar product of two general elements of the vector space is an infinite
sum. The product will be regarded as formal and questions of convergence will
be postponed for the time being.

The reversal of strings will be used to define conjugation denoted with
a dagger. Then for any base string C.
\begin{equation}
		    C^\dagger \bullet C = -1
\end{equation}
The conjugate of a general element of the space is defined by reversing each
base element and in the case of the complex space the complex conjugate of the
components is also taken. I.e.
\begin{equation}
                     \Phi^\dagger = \sum \Phi_C^* C^\dagger
\end {equation}
To add the structure of an associative algebra the product $AB$ of two strings
in the space is defined by joining them when the end of $A$ matches the
beginning of $B$ reversed. It is necessary to add all the ways in which
this can be done e.g.
\begin{equation}
        (1,4,3,1,7)(7,1,5,1) = (1,4,3,5,1) + (1,4,3,1,1,5,1)
\end{equation}
In the case where the last point of the first string is not the first
point of the second the product is always zero. This ensures that
string models are local, i.e. strings which do not intersect should
not interact directly. E.g.
\begin{equation}
                     (1,4,3,1,7)(1,6) = 0
\end{equation}
Also if the whole of one
of the strings in a product matches, the last event is not cancelled e.g.
\begin{equation}
                     (1,4,3,1,7)(7,1) = (1,4,3,1,1)
\end{equation}
It can now be checked that these rules define an associative multiplication,
\begin{equation}
                     A(BC) = (AB)C
\end{equation}
and that it is closed over strings of length 2 and greater.
There is the usual relation between conjugation and multiplication
\begin{equation}
                     (AB)^\dagger = B^\dagger A^\dagger
\end{equation}
Given such an associative algebra an infinite dimensional Lie algebra can be
defined with the Lie product given by the anticommutator,
\begin{equation}
                     A \wedge B = AB - BA
\end{equation}
Here are some examples to clarify special cases,
\begin{equation}
  (1,4,3,1,7) \wedge (7,1,3,2) = (1,4,2) + (1,4,3,3,2) + (1,4,3,1,1,3,2)\\
  (1,4,3,1,7) \wedge (1,6,6) = 0\\
  (1,4,3,1,7) \wedge (7,5,4,1) = (1,4,3,1,5,4,1) -
                     (7,5,3,1,7) - (7,5,4,4,3,1,7)\\
  (1,4,3,1,7) \wedge (7,1,3) = (1,4,3,3) + (1,4,3,1,1,3)\\
  (3,1,7) \wedge (7,1,3) = (3,1,1,3) + (3,3) - (7,1,1,7) - (7,7)\\
\end{equation}
This product satisfies the Jacobi identity,
\begin{equation}
   A \wedge (B \wedge C) + B \wedge (C \wedge A) + C \wedge (A \wedge B) = 0
\end{equation}
and is also a differential operator on the associative algebra in the sense
that
\begin{equation}
	(AB) \wedge C = (A \wedge C)B + A(B \wedge C)
\end{equation}
With the wedge product the algebra is an infinite dimensional Lie Algebra
and it
defines a group by exponentiation. Using the structure constants for the
algebra the Lie product can be written
\begin{equation}
    A \wedge B    =   \sum f^C_{AB} C
\end{equation}
string indices are formally lowered and raised by reversing the direction of
the string so that
\begin{equation}
                       f_{CAB} = f^{C^\dagger}_{AB}
\end{equation}
The three strings C, A and B are said to form a triplet is $f_{ABC}$ is
plus one, and an anti-triplet if it is minus one. They are a triplet
if and only if they are all different and the end of A matches the beginning
of B, the end of B matches the beginning
of C and the end of C matches the beginning of A without any events being
left out of used twice. Anti-triplets are triplets with two of the strings
interchanged. It follows that the structure constants are fully antisymmetric.
\begin{equation}
         f_{ABC} = f_{BCA} = f_{CAB} = -f_{ACB} = -f_{CBA} = -f_{BAC}
\end{equation}
The following important relation is also valid
\begin{equation}
   A \bullet ( B \wedge C ) = (A \wedge B) \bullet C  = -f_{ABC}
\end{equation}
To define an event symmetric model with this discrete string
symmetry take field variables $\Psi$ from the fundamental representation.
In the basis used to define the group the components of $\Psi$ can be
regarded as a family of tensor fields and we can write,
\begin{equation}
   \Psi = \psi_{ab} (a, b) + \psi_{abc} (a, b, c) + ...\ldots
\end{equation}
The infinitesimal transformations are generated by an element $\epsilon$ of
the algebra as follows,
\begin{equation}
          \delta \Psi = \Psi \wedge \epsilon
\end{equation}
Invariant actions can be constructed using the ``vector'' and ``scalar''
products,
\begin{equation}
   S  =  m \Psi \bullet \Psi  +
   \beta (\Psi \wedge \Psi) \bullet (\Psi \wedge \Psi)
\end{equation}
Here the $\Psi$ must be an anti-commuting field since otherwise the vector
product would be zero. The Model has an infinite number of field variables
even on a finite number of events.

There are other invariants which can be defined by generalising the definition
of trace for matrix models.
Examining the string algebra it is observed that there is a sub-algebra
spanned by the rank two bases,
\begin{equation}
  (a, b) (c, d) = \delta_{bc} (a, d) \\
  (a, b) \wedge (c, d) = \delta_{bc} (a, d) -
                                       \delta_{ad} (c, b)
\end{equation}
The sub-algebra is isomorphic to multiplication of $N \times N$ Real or Complex
matrices and the corresponding Lie sub-algebra is isomorphic to the Lie algebra
of the general linear group $gl(N,{\cal R})$ or $gl(N,{\cal C})$. The
generalised trace is the sum over even rank components of the vector space
whose basis elements are palindromic sequences. e.g.
\begin{equation}
                   Tr(1,4,4,1) = Tr(3,3) = 1\\
                   Tr(2,3) = Tr(1,2,1) = 0
\end{equation}
It can be verified that the trace has the familiar properties,
\begin{equation}
                   Tr(AB) = Tr(BA)\\
                   Tr(A \wedge B) = 0
\end{equation}
It follows that the trace is an invariant of the group and that the
traceless elements of the algebra form a Lie sub-algebra which contains
the Lie algebra of the special linear group $sl(N,{\cal R})$ or
$sl(N,{\cal C})$. It is also possible to form a vector of invariants from the
palindromic strings of odd length.

There are now many invariants which can be formed in a fashion analogous to
those for matrix models for example the following invariant might be
interesting as an action,
\begin{equation}
   S  =  m Tr(\Phi^2) + \beta Tr(\Phi^4)
\end{equation}
where \Phi is a representation with commuting components.

This action presents an immediate problem: It is not positive definite.
That is not surprising since analogous matrix models based on the general
linear group would also have problems over
positive definitiveness of the action. Evidently we must find the string groups
which are extensions of the compact groups such as $SO(N)$ in the same way as
this group is a string extension to $GL(N)$. This is not too difficult. The
elements of the group which are the negative of their conjugate form another
Lie-subalgebra because of the general identity
\begin{equation}
                 ( A \wedge B )^\dagger = - A^\dagger \wedge B^\dagger
\end{equation}
These algebras can be regarded as the string extensions of $so(N)$ in the real
case and $u(N)$ in the complex case. There is still a problem with positivity
of actions defined using the trace invariant for these groups. The trace square
$Tr(\Phi^2)$ for example contains such non-square terms as,
\begin{equation}
                 \phi_{ab}\phi_{bdda}
\end{equation}
A positive definite action can be defined using the scalar product.
\begin{equation}
   S  =  m \Phi \bullet \Phi + \beta \Phi^2 \bullet \Phi^2
\end{equation}
This action and other similar actions define statistical or quantum models
of event symmetric string theory. It is important to recognise that the model
has  an infinite number of
degrees of freedom even for finite $N$. It would be necessary to demonstrate
that it can give a well defined model despite this.

\subsection{Supersymmetric String Groups}

An attractive feature of the discrete string groups on event symmetric
space-time is that supersymmetric versions can be constructed in a very natural
way.

The open string extension of $GL(N)$ can be made supersymmetric in a
straightforward fashion. The Infinite dimensional algebra is modified
to a graded Lie algebra by identifying the base strings of odd length with odd
elements and those of even length with even elements. The Lie product of two
odd base elements is then made symmetric instead of anti-symmetric. I.e.
\begin{equation}
                     A \wedge B = AB - (-1)^{par(A)par(B)} BA
\end{equation}
where $par(A)$ is the parity of A, 0 if it is of even length and 1 if it is
of odd length.

The components of an element in the representation of the algebra can be taken
to be commuting for even base elements and anti-commuting for odd base
elements.
\begin{equation}
                      \Phi = \sum \Phi_C C\\
                \Phi_A \Phi_B = (-1)^{par(A)par(B)} \Phi_B \Phi_A
\end{equation}
Then the lie product for elements of the representation will be anticommuting.
\begin{equation}
                     \Phi \wedge \Psi = - \Psi \wedge \Phi
\end{equation}
The adjoint operator must fulfill the usual relation
\begin{equation}
                     (\Phi \Psi)^\dagger = \Psi^\dagger \Phi^\dagger
\end{equation}
This is achieved by modifying the previous definition to include a factor of
$i$ when taking the adjoint of an odd element. This restricts us to the
complex version of the model.
\begin{equation}
                     \Phi^{\dagger} = \sum i^{par(C)} \Phi_C^* C^{\dagger}
\end{equation}
The rest is exactly the same as for the non-supersymmetric case. A
representation of a reduced Lie sub-algebra is defined as those elements which
satisfy,
\begin{equation}
                     \Phi^\dagger = -\Phi
\end{equation}
Actions for a model based on this representation are also the same as
before. In general the action may contain any powers in the algebra squared
with the scalar product.
\begin{equation}
   S  =  g_1 \Phi \bullet \Phi + g_2 \Phi^2 \bullet \Phi^2 +
                                 g_3 \Phi^3 \bullet \Phi^3 + \ldots
\end{equation}

This supersymmetric generalisation is an analogue of the supersymmetric
generalisation of matrix models already described. A positive definite
action can again be defined on the fundamental representation. This model
may prove to be one of the most important models on Event Symmetric
space-time because of its high symmetry and its obvious relevance to
super-string theories.

It is possible that interesting physics exists in these models in
a large $N$ double scaling limit with the constants $g_i$ scaled as
functions of $N$.

\section{Conclusions}

We started from a simple principle that physics could be described by
an event symmetric model and considered open string field theory on
event-symmetric space-time as a possibility. The models which result
unify space-time symmetries and string gauge groups in a simple elegant
way. Furthermore they can be recognised as natural extensions of random
matrix models which are known to be of interest in the non-perturbative
study of string theories.

It is possible that techniques used to study matrix models in the
double scaling limit may also be applicable to event-symmetric string
theories and that their study may provide further insight towards understanding
the nature of superstring theories and of space-time.

More optimistically it is possible that event symmetric string models
are pre-theories which induce continuum string theories through a
process of spontaneous symmetry breaking of the event symmetry. It is
possible that string theories restore their full symmetry including
the event symmetry beyond their Hagedorn temperature.

The author welcomes all comments and corrections
by e-mail to phil@galilee.eurocontrol.fr


\section*{References}
\begin{references}


\bibitem{Planck(1899)}, Planck {\sl Ueber irreversible Strahlungsvorgaenge,
         Sitzungsber}. Deut Akad Wiss Berlin, {\bf Kl}.
	 Math-Phys Tech, 440-480 (1899).

\bibitem{Hagedorn(1968)} R. Hagedorn, {\sl Hadronic Matter Near the Boiling
         Point}, Nuovo Cim {\bf 56A}, 1027 (1968)

\bibitem{Wess,Zumino(1974)} J. Wess, B. Zumino, {\sl Supergauge
         transformations in four dimensions}, Nucl Phys {\bf B70}, 39 (1974)

\bibitem{Green,Schwarz(1981a)} M.B. Green, J.H. Schwarz, {\sl Supersymmetrical
         Dual String Theory}, Nucl Phys {\bf B181}, 502 (1981)

\bibitem{Green,Schwarz(1981b)} M.B. Green, J.H. Schwarz, {\sl Supersymmetrical
         Dual String Theory (II), Vertices and trees},
	 Nucl Phys {\bf B198}, 252 (1981)

\bibitem{Green,Schwarz(1982)} M.B. Green, J.H. Schwarz, {\sl Supersymmetrical
         String Theories}, Phys Lett {\bf 109B}, 502 (1982)

\bibitem{Wheeler(1984)} J. A. Wheeler in {\it Problems in Theoretical Physics},
         ed. A. Giovannini, F. Mancini, and M. Marinaro (University of Salerno
 	 Press, Salerno, 1984)

\bibitem{Witten(1986)} E. Witten, { \sl Noncommutative geometry and
         string field theory }, Nucl Phys {\bf B268} 253 (1986).

\bibitem{Gross,Mende(1987)} D. Gross, P.F. Mende, {\sl The high energy
         behaviour of string scattering amplitudes},
         Phys Lett {\sl B197}, 129 (1987)

\bibitem{Gross,Mende(1988)} D. Gross, P.F. Mende, {\sl String theory beyond
         the Planck scale}, Nucl Phys {\bf B303}, 407 (1988)

\bibitem{Gross(1988)} D. Gross, Phys Rev Lett {\bf 60}, 1229 (1988)

\bibitem{Gross,Perival(1988)} D. Gross, V. Perival, Phys Rev Lett {\bf 60},
         2105 (1988)

\bibitem{Kaku(1988)} M. Kaku, {\sl Introduction to Superstings},
         Springer-Verlag

\bibitem{Atick,Witten(1988)} J.J. Atick, E. Witten, {\sl The Hagedorn
         Transition and the Number of Degrees of Freedom of String Theory},
	 Nucl Phys {\bf B310}, 291 (1988)

\bibitem{Kaku(1991)} M. Kaku, {\sl Strings, Conformal Fields and Topology},
         Springer-Verlag

\bibitem{Zwiebach(1992)} B. Zwiebach, {\sl Closed String Field Theory:
         Quantum Action and the B-V Master Equation },IASSNS-HEP-92/41,
	 MIT-CTP-2102, hep-th/9206084

\bibitem{Antoniadis,Ferrara,Kounnas(1994)} I. Antoniadis, S. Ferrara,
         C. Kounnas, {\sl Exact Supersymmetric String Solutions in Curved
	 Gravitational Backgrounds}, CERN-TH.7148/94, UCLA/94/TEP/5,
	 hep-th/9402073

\bibitem{Gibbs(1994)} P.E. Gibbs, {\sl Models on Event Symmetric Space-Time},
         PEG-01-94, hep-th/9404139

\end{references}
\end{document}